\documentstyle[12pt,amsthm,draft]{article}
\def\@aabuffer{}

\def\be{\begin{equation}}
\def\ee{\end{equation}}
\def\bea{\begin{eqnarray}}
\def\eea{\end{eqnarray}}

\def\bbbr{\mathbf R}
\def\aa{\alpha}

\def\KM{{\rm  KM}}

\def\gr{g^{\rm RL}}
\def\gs{g^{\rm SL}}
\def\Ri{Riemannian\ }

\def\Ker{{\rm Ker}\,}
\def\Conj{\bar}

\def\Diag{{\bf Diag}}
\def\T{T}
\def\iM{{\cal M}}
\def\iMR{\iM^{\prime}}

\def\t{{\rm Tr}\,}
\def\Im{{\rm Im}\,}
\def\Re{{\rm Re}\,}

\def\fel{\textstyle{1 \over 2}}
\def\t{{\rm Tr}\,}
\def\im{{\rm i}}

\def\<{\langle}
\def\>{\rangle}
\def\iS{{\cal S}}

\def\iM{{\cal M}}

\def\TT{{\bf T}}

\def\Diag{{\bf Diag}}

\def\pard{\partial}

\def\sq{\hbox{\rlap{$\sqcap$}$\sqcup$}}
\def\qed{\ifmmode\sq\else{\unskip\nobreak\hfil
\penalty50\hskip1em\null\nobreak\hfil\sq
\parfillskip=0pt\finalhyphendemerits=0\endgraf}\fi}

\def\eqref#1{(\ref{#1})}
\def\pure{\cal P}

\newcommand{\Lift}[1]{\widehat {#1}}

\newtheorem{theorem}{Theorem}[section]
\newtheorem*{theorem*}{Theorem}

\theoremstyle{definition}

\newtheorem*{definition*}{Definition}

\begin{document}
\title{Extending the Fisher metric to density matrices \thanks{Published
in {\it Geometry in Present Days Science}, eds. O.E. Barndorff-Nielsen and 
E.B. Vendel Jensen, 21--34 (World Scientific, 1999), written version of
the conference talk at Aarhus University in 1997.}}

\author{D. Petz and Cs. Sud{\'a}r \\ \phantom{mmmmmmm} \\ 
Department for Mathematical Analysis,\\ Technical University of
Budapest, H-1521 Budapest, Hungary}
\maketitle
\begin{abstract}
Chentsov studied Riemannian metrics on the set of
probability measures from the point of view of decision theory.
He proved that up to a constant factor the Fisher information is
the only metric which is monotone under stochastic transformation.
The present paper deals with monotone metrics on the space of
finite density matrices on the basis of motivation provided by
quantum mechanics. A characterization of those metrics
is given in terms of operator monotone functions. Several
concrete metrics are constructed and analyzed, in particular,
instead of uniqueness in the probabilistic case, there is a large
class of monotone metrics, some of which appeared long time ago
in the physics literature. Moreover a limiting procedure to pure
states is discussed.
\end{abstract}

\section{Introduction}
The idea of statistical distance between two probability
distributions goes back to Fisher who was interested in a
quantity which shows how difficult it is to decide between
two probability measures by statistical sampling. He found
that the spherical representation of the probability
simplex is adequate. The probability distributions
$(p_1,p_2,\dots, p_n)$ on $n$ points form an
$(n-1)$-dimensional simplex $\iS_{n-1}$, since $p_i \ge 0$ and
$\sum_i p_i=1$. If we introduce the parameters $z_i=2\sqrt{p_i}$,
then $\sum_i z_i^2=4$ and the probability simplex is para\-met\-rized
as a portion of the $n$-sphere. Let $z(t)$ be a curve on the
sphere. The square of the length of the tangent is
\begin{equation} \label{eq:1.2}
\< \pard_t z, \pard _t z\>=\sum_i (\pard_t z_i)^2=\sum_i p_i(t)
(\pard_t \log p_i(t))^2\,,
\end{equation}
which is the Fisher information. The geodesic distance between
two probability distributions $Q$ and $R$ can be computed along a
great circle and it is
\[
d(Q,R)=2 \arccos \sum_{i=1}^n \sqrt{p_ir_i}\,.
\]
One observes that the geodesic distance is a simple transform of the
Hellinger distance. Namely,
\[
d_H(Q,R)\equiv \sqrt{\textstyle{\sum_{i=1}^n (p_i^{1/2}-r_i^{1/2})^2}}=2
\sin \big( d(Q,R)/4\big) \, .
\]
In applications of mathematical statistics one often meets a family of
distributions parametrized by a real number or more generally by
$\theta \in \bbbr^m$. An example is the family $N(\mu,\sigma)$ of
normal distributions with mean $\mu \in \bbbr$ and variance
$\sigma \in \bbbr^+$. An $n$-tuple $(\xi_1,\xi_2,\dots ,
\allowbreak \xi_n)$ of random variables is called an unbiased
estimator of the parameter $\theta$ if $E(\xi_i)=\theta_i$ for $1 \le i \le n$. In statistical
problems an unbiased estimator can be used to estimate the true
value of the parameter $\theta$ on the basis of a sample. The
variance of the estimator is desired to be small in order to have
an effective estimation. The classical Cram{\'e}r-Rao inequality
is related to that point. The $m\times m$ covariance matrix
$E(\xi_i\xi_j)-E(\xi_i)E(\xi_j)$  is always larger than the
inverse of the Fisher information matrix. The latter is
independent of the estimator $(\xi_1,\xi_2,\dots ,\xi_n)$
and one desirable property of an unbiased estimator is
closeness of the covariance matrix to the inverse Fisher
information matrix.

In quantum mechanics, the state space of an
$n$ level system is identified with the set of all
$n\times n$ positive semidefinite complex matrices of
trace 1, they are the so-called density matrices. Let
$\iM_n$ stand for the set of all positive definite density
matrices. We can parametrize $D=(D_{ij})\in \iM_n$ by the
real numbers $\Re D_{ij}$, $\Im D_{ij}$ $(1\le i <j \le n)$ and by the positive numbers $D_{ii}$ $(1 \le i
\le n-1)$. In this way $\iM_n$ may be embedded into the Euclidean
$k$-space with $k=n^2-1$ and becomes a manifold. At each point $D\in
\iM_n$ the tangent space $\T_D(\iM_n)$ is identified with the set
of all traceless selfadjoint matrices. One observes that the
probability simplex is embedded into $\iM_n$, since every
probability distribution on the $n$-point space gives a diagonal
density matrix in the obvious way:
\[
\iS_{n-1} \ni (p_1,p_2,\dots, p_n)\mapsto \Diag
(p_1,p_2,\dots, p_n) \in \iM_n.
\]

The aim of the present paper is a search for possible Riemannian
metrics on the space of density matrices of a finite dimensional
space. Without some restrictions this would be pointless, the emphasis
is put on statisticaly relevant metrics which on the submanifold of
probability distributions recover the Fisher information metric.

\section{Chentsov's approach to the problem}

Chentsov was led by decision theory when he considered a category whose
objects are probability spaces and whose morphisms are Markov
kernels. Although he worked in \cite{Cencov-1982} with arbitrary
probability spaces, his idea can be demonstrated very well on
finite ones. In this case a Markov kernel from the probability
$(n-1)$-simplex $\iS_{n-1}$ to an $(m-1)$-simplex
$\iS_{m-1}$ is an $m\times n$ stochastic matrix. If $\Pi$ is such
a matrix
and $P\in \iS_n$ then $\Pi P \in \iS_m$ is considered more random than
$P$. If we want to represent probability distributions as column
vectors then the matrix $\Pi$ has to be column-stochastic, that is,
$\sum_i \Pi_{ij}=1$ for every $j$. An example of randomization
comes from identification of two outcomes of our random
experiment. This is described by a 0-1 matrix with one 1 in each
row except for one where two 1's stand. In statistical physical literature
the term coarse graining is more often used than randomization but they
stand for the same concept.

Generally speaking, the parametrized family $(Q_i)$ is more random than
the parametrized family $(P_i)$ (with the same parameter set) if there
exists a stochastic matrix $\Pi$ such that $\Pi P_i =Q_i$ for every
value of the parameter $i$. Two parametric families  $(P_i)$  and
$(Q_i)$ are equivalent in the theory of statistical inference if there
are two stochastic matrices $\Pi^{(12)}$  and $\Pi^{(21)}$ such that
\begin{equation} \label{eq:2.1}
\Pi^{(12)} P_i =Q_i \quad {\rm and}\quad \Pi^{(21)} Q_i =P_i
\end{equation}
for every $i$. Chentsov defined a numerical function $f$ given
on pairs of measures to be invariant if
\begin{equation} \label{eq:2.2}
(P_1,P_2)\sim(Q_1,Q_2)\quad {\rm implies}\quad f(P_1,P_2)=f(Q_1,Q_2)
\end{equation}
and monotone if
\begin{equation} \label{eq:2.3}
f(P_1,P_2) \ge f(\Pi P_1, \Pi P_2)\,.
\end{equation}
for every stochastic matrix $\Pi$. A monotone function $f$ is obviously
invariant. Statistics and information theory know a lot of monotone
functions, relative entropy
\begin{equation} \label{eq:2.4}
S(P,Q)=\sum_i p_i (\log p_i -\log q_j )
\end{equation}
and its generalizations. If a Riemannian metric is given on
all probability simplexes, then this family of metrics is
called invariant (respectively, monotone) if the corresponding
geodesic distance is an invariant (respectively, monotone)
function. Chentsov's greate achievement was to show that up to a
constant factor the Fisher information \eqref{eq:1.2} yields
the only monotone family of Riemannian
metrics on the class of finite probability simplexes (\cite{Cencov-1982}).

A decade later Chentsov turned to the quantum case, where the
probability simplex is replaced by the set of density matrices. A linear
mapping between two matrix spaces sends a density matrix into a density
if the mapping preserves trace and positivity (i.e., positive
semidefinitness). By now it is well-understood that completely positivity
is a natural and important requirement in the quantum case.
Therefore, we call a trace preserving completely positive mapping
stochastic. One of the equivalent forms of the completely positivity of
a map $T$ is the following.
\[
\sum_{i=1}^n  \sum_{j=1}^n a_i^*T(b_i^*b_j)a_j \ge 0
\]
for all possible choice of $a_i$, $b_i$ and $n$. A completely positive
mapping $T$ satisfies the Schwarz inequality: $T(a^*a)\ge
T(a)^*T(a)$.

Chentsov recognized that stochastic mappings are the appropriate
morphisms in the category of quantum state spaces. (The monograph
\cite{Alberti-Uhlmann-1981}
contains more information about stochastic mappings, see also
\cite{Kraus-1983}.) The
above definitions of invariance and monotonicity make sense when
stochastic matrices are replaced by stochastic mappings. Chentsov (with
Morozova) aimed to find the invariant (or monotone) Riemannian metrics
in the quantum setting as well. They obtained the following
result
(\cite{Morozova-Chentsov-1990}).
Assume that a family of Riemannian metrics is given on all spaces of
density matrices which is invariant, then there exist a function
$c(x,y)$ and a constant $C$ such that the squared length of a tangent
vector $A=(A_{ij})$ at a diagonal point
$D=\Diag(p_1,p_2,\dots,p_n)$ is of the form
\begin{equation} \label{eq:2.5}
C\sum_{k=1}^n p_{k}^{-1} A_{kk}^2+2\sum_{j < k}
c(p_j,p_k)|A_{jk}|^2\,.
\end{equation}
Furthermore, the function $c(x,y)$ is symmetric and
$c(\lambda x,\lambda y)=\lambda^{-1}c(x,y)$.
This result of Morozova and Chentsov was not complete. Although they had
proposals for the function $c(x,y)$, they did not prove monotonicity or
invariance of any of the corresponding metrics. A complete result was obtained
in \cite{Petz-JMATP4-1994} and \cite{Petz-LINAA2} but before presenting
it here we make  a few comments on \eqref{eq:2.5}.

Both the function $c(x,y)$ and the constant $C$ are independent
of the matrix size $n$. Restricting ourselves to diagonal matrices, which is in
some sense a step back to the probability simplex, we can see that
there is no ambiguity of the metric. Loosely speaking, the
uniqueness
result of the simplex case survives along the diagonal and the
offdiagonal provides new possibilities for the definition of a
stochastically invariant metric on the space $\iM$ of invertible
density matrices. In other words, the tangent
space $T_D (\iM)$ at $D$ decomposes as
\begin{equation} \label{eq:2.6}
T_D (\iM)= T_D (\iM)^{c}\oplus T_D (\iM)^{o}\, ,
\end{equation}
where $T_D (\iM)^{c}=\{ A\in T_D (\iM)  : [A,D]=0\}$ and $
T_D (\iM)^{o}$ is the orthogonal complement of $T_D (\iM)^{c}$
with respect to the Hilbert-Schmidt inner product of matrices.
The monotone metric is unique on $T_D (\iM)^{c}$,
\begin{equation} \label{eq:2.7}
K_D (A,A)= C \t D^{-1}A^2 \quad {\rm if}\quad  A\in T_D (\iM)^{c}
\end{equation}
and the function $c(x,y)$ determines the metric on the orthogonal
complement.

If a distance between density matrices expresses statistical
distinguishability then this distance must decrease under
coarse-graining. A good example of coarse-graining arises when a density
matrix is partitioned in the form of a $2\times 2$ block matrix, and the
coarse-graining forgets about the offdiagonal:
\[
\left( \matrix{A & B \cr B^* & C }\right)
\quad \longmapsto
\left( \matrix{A & 0 \cr 0 & C }\right)
\]
In the mathematical formulation, a coarse-graining is a completely
positive mapping which preserves the trace and hence sends
density matrix into density matrix. Such mapping will be called
stochastic below. A Riemannian metric is defined to be monotone if the
differential of any stochastic mapping is a contraction (in the
sence that it is norm decreasing). If the affine
parametrization is considered, then $D_t=D+tA$ is a curve for an
invertible density $D$ and for a selfadjoint traceless $A$. Under a
stochastic mapping $\TT$ this curve is transformed into
$\TT(D_t)=\TT(D)+t\TT(A)$
provided that $\TT(D)$ is an invertible density and the real number $t$
is small enough.
The monotonicity condition for the Riemannian metric $g$ on $\iM_n$
reads as
\begin{equation} \label{eq:2.8}
g_{\TT(D)}\big(\TT(A),\TT(A)\big) \le g_D(A,A)\,,
\end{equation}
for any invertible density $D$, for any traceless selfadjoint
matrix $A$ and for any stochastic mapping $\TT$.
Our goal is to show many examples of monotone metrics and to give their
characterization in terms of operator monotone functions.

\section{Monotone metrics}

Let us recall that a function $f:\bbbr^+ \to \bbbr$ is called operator
monotone if the relation $0\le K \le H$ implies $0\le f(K)\le f(H)$ for
any matrices $K$ and $H$ (of any order). The theory of operator monotone
functions was established in the 1930's by L{\"o}wner and there are
several reviews on the subject, for example \cite{Ando-1979}, \cite{Hansen-Pedersen-1982} are
suggested.

The following result was obtained in \cite{Petz-LINAA2}.

\begin{theorem}
There exists a one-to-one correspondence between
monotone metrics and operator monotone functions $f:\bbbr^+ \to
\bbbr^+$ such that $f(t)=tf(t^{-1})$. If
$D=\Diag(p_1,p_2,\dots,p_n)$, then the metric
corresponding to $f$ is of the form
\begin{equation} \label{eq:3.1}
\sum_{j=1}^n \sum_{k=1}^n c(p_j,p_k)|A_{jk}|^2\, ,
\end{equation}
where $c(x,y)=1/y f(x/y)$.
\end{theorem}
The proof of this result is given in the original paper. Here we
remark that the metric \eqref{eq:2.5} can be written by means of a
certain function $f$ such that $c(x,y)=1/y f(x/y)$ holds. The point is
to demonstrate, on the one hand that this function $f$ must be
operator monotone and, on the other hand that every operator
monotone
function provides a monotone metric. The symmetry condition
$f(t)=tf(t^{-1})$ is equivalent to the condition that the
Riemannian inner product is real valued on the selfadjoint
tangent vectors. It seems natural to normalize metrics such a way
that on the submanifold of diagonal matrices the standard Fisher
metric should appear. In this case one can say following
Uhlmann that the metric is Fisher adjusted. This normalization is
equivalent to
the condition $f(1)=1$. Below we always assume that
$f(1)=1$, that is, we restrict our discussion to Fisher adjusted
metrics. Some examples of functions $f$ satisfying the hypothesis of
Theorem 3.1 are the following.
\begin{equation} \label{eq:3.2}
\frac{ 2x^{\alpha+1/2}}{ 1+x^{2\alpha}},
\quad \frac{x-1 }{ \log x},
\quad \frac{x-1 }{ \log x}\,\frac{2\sqrt{x} }{ 1+x},
\quad \Big(\frac{x-1 }{ \log x}\Big)^2\,\frac{2 }{ 1+x}, \quad
\frac{1+x }{ 2}
\end{equation}
where $0 \le \alpha \le 1/2$.

It is worthwhile to note that Kubo and Ando established a
correspondence between operator monotone functions and
means of positive operators. Our condition $f(t)=tf(t^{-1})$ on
the operator monotone function $f$ is equivalent to the symmetry
of the corresponding operator mean. The smallest mean is the
harmonic one. This corresponds to the function $f(t)=2t/(t+1)$
and gives the metric
\begin{equation} \label{eq:3.3}
\gr_D (A,B) = \fel \t D^{-1}(AB+BA).
\end{equation}
Since a larger function $f$ yields a smaller metric, we have

\begin{theorem}
{The Riemannian metric \eqref{eq:3.3} is monotone and it is
the largest among all Fisher-adjusted monotone metrics.}
\end{theorem}
One can see monotonicity of \eqref{eq:3.3} directly. The operator
inequality
\begin{equation} \label{eq:3.4}
\TT (K)\TT (D)^{-1}\TT(K)^*\le \TT (KD^{-1}K^*)
\, ,
\end{equation}
holds for positive invertible $D$ for every stochastic
mapping \cite{Choi-1980}, \cite{Lieb-Ruskai-1974}. Taking the trace of
both sides of \eqref{eq:3.4}, we conclude monotonicity.

The arithmetic operator mean is the largest symmetric mean and
it gives the smallest metric which is usually called the metric
of the symmetric logarithmic derivative.

\begin{theorem}
Among all Fisher-adjusted monotone metrics the smallest one
is given as
\begin{equation} \label{eq:3.5}
\gs_D (A,B) = \t AG,
\end{equation}
where $G$ is the unique solution of the equation
\begin{equation} \label{eq:3.6}
DG+GD=2B.
\end{equation}
\end{theorem}
The metrics $\gr$ and $\gs$ appeared in connection with
generalizations of the Cram\'er-Rao inequality and $\gs$ play
important role in the work of Uhlmann when he extends Berry phase
to mixed states from the pure ones. Is is rather instructive to
have a look at the simple $2\times 2$ case.

Dealing with $2\times 2$ density matrices, we conveniently use
the so-called Stokes parametrization.
\begin{equation} \label{eq:Stokes}
D_x = \fel (I + x_1\sigma_1+ x_2\sigma_2+ x_3 \sigma_3)\equiv
\fel (I+x\cdot \sigma)
\end{equation}
where $\sigma_1, \sigma_2, \sigma_3$ are the Pauli matrices and
$(x_1, x_2, x_3) \in \bbbr^3$ with $x^2_1 + x^2_2 + x^2_3 \le 1$.
The monotone metrics on $\iM_2$ are rotation invariant in the
sense that they depend only on $r=\sqrt{x^2+y^2+z^2}$ and split into
radial and tangential components as follows.
\begin{equation} \label{eq:3.7}
ds^2={1 \over 1-r^2}dr^2+{1 \over 1+r}g\Big({1-r \over 1+r}\Big)dn^2
\quad {\rm where}\quad g(t)={1 \over f(t)}\,.
\end{equation}
The radial component is independent of the function $f$. In
case of $f(t)=(t+1)/2$ we have constant tangential component.
In the case of $f(t)=2t/(1+t), ds^2=(1-r^2)^{-1}(dr^2+dn^2).$
Hence both the smallest and the largest metrics possess a
rather particular form.

The  limit of the tangential component exists when $r \to 1$ if
$f(0)\ne 0$. In this way the standard metric is obtained on the set
of pure states, up to a constant factor. In case of larger density matrices,
pure states form a small part of the topological boundary of the invertible
density matrices. Hence, in order to speak about the extension of a Riemannian
metric on invertible densities to pure states, a rigorous meaning of the
extension should be given. This is the subject of the paper
\cite{Sudar-1996} and will be discussed in the next section.

It is remarkable that quantum statistical mechanics seems to prefer
another metric, different from the smallest and from the largest
one. This is termed the Kubo-Mori, or Boguliubov metric and
sometimes canonical correlation. In the above used
affine parametrization of the state space the Kubo-Mori metric takes
the form
\[
g_D^{\KM}(A,B)=\int_0^\infty \t (D+t)^{-1} A (D+t)^{-1}B\,
dt\,.
\]
In order to see that this is the usual Kubo-Mori inner product,
we rewrite it in the logarithmic coordinate system instead of
the affine one. In terms of the inverse Kubo transforms
\begin{eqnarray}
A'&=&\int_0^\infty (D+s)^{-1}A(D+s)^{-1}\,ds, \\
B'&=&\int_0^\infty (D+s)^{-1}B(D+s)^{-1}\,ds
\end{eqnarray}
we have
\begin{equation} \label{eq:3.8}
g_D^{\KM}(A,B)=\int_0^1 D^t A'D^{1-t}B'\, dt\,.
\end{equation}

\begin{theorem}
Assume that a Fisher adjusted monotone metric $g$ is obtained
from a smooth  function $G:\bbbr^+\to \bbbr$ by
\[
g(A,B)(D)={\pard \over \pard t\pard s}\Big|_{t=s=0}
\t G(D+tA+sB)\,.
\]
Then $g(A,B)$ is the Kubo-Mori inner product.
\end{theorem}

\begin{proof}
When $A,B$ and $D$ commute, we have
\[
{\pard \over \pard t\pard s}\Big|_{t=s=0}
\t G(D+tA+sB)=\t G''(D)AB\,.
\]
Since we assumed that the metric is Fisher-adjusted,
$G''(t)=t^{-1}$ and we have $G(t)=t\log t+Ct+D$ and
the differentiation gives the Kubo-Mori metric.
\end{proof}

The above proof also gives that the Kubo-Mori metric is the
negative Hessian of the von Neumann entropy functional on the
state space. Recall that the von Neumann entropy is the
Boltzmann-Shannon entropy of the eigenvalues, that is,
\[
S(D):= -\t (D\log D)\, .
\]
Differentiation of entropy-like functional is a good method to
obtain monotone metrics. In one variable Theorem 3.4 doest not
allow
many possibilities but in the two variable case one can get
more metrics. A typical two-variable-entropy is the relative
entropy $\t(D_1 (\log D_1 -\log D_2))$ which is a member of the family
of $\aa$-entropies. If $-2 <\aa< 2$, then
\begin{equation}\label{eq:3.21}
S_\aa (D_1,D_2 )= {4 \over 1-\aa^2}
\t (I-D_2^{{1+\aa \over 2}}D_1^{-{1+\aa \over 2}})D_1
\end{equation}
is jointly convex. The metric
\begin{equation}\label{eq:3.22}
{\partial^2 \over \partial t \partial u}
S_\aa(D+tA,D+uB)\Big\vert_{t=u=0}=K_D^{\aa}(A,B)
\end{equation}
was studied first by Hasegawa \cite{Hase-1993}, \cite{Hase-1995}
and its monotonicity was proved in \cite{Hase-Petz-1996} and
\cite{Hase-Petz-1997}. Note that the limit $\aa \to \pm 1$ in the
formulas recovers the usual relative entropy and the Kubo-Mori
metric.
Since (22) is a monotone metric, it is really interesting on
tangent vectors orthogonal to the commutator of $D$:
\begin{equation}\label{eq:3.23}
K_D^{\aa}(\im  [D,X],\im[D,X])={ 2 \over 1-\aa^2} \t
\big([D^{1-\aa \over 2}, X][D^{1+\aa \over 2},X]\big)\,,
\end{equation}
where $X$ is selfadjoint. It is worthwile to point out the similarity
to the skew information proposed by Wigner, Ya\-na\-se and Dyson (apart from
a constant factor), see \cite{Wigner-Yanase-1963} or p. 49 in
\cite{Ohya-Petz-1993}. The operator monotone functions
corresponding to (22) are
\[
f_\aa(x)={\beta(1-\beta)}\,{ (x-1)^2 \over (x^\beta -1)
(x^{1-\beta}-1)}\, .
\]
where $\beta=(1-\aa)/2$.

The following characterization of the $\aa$-metrics was obtained
in \cite{Hase-Petz-1997}.

\begin{theorem}
In the class of symmetric monotone metrics, the Wigner-Ya\-na\-se-Dyson skew
information (i.e.the $\alpha$-metric (22)) is characterized by
the property that
\[
K_{\rho}(A,B) = \frac{\partial ^2}{\partial t\partial s} {\rm Tr}
g(\rho + tA)g^*(\rho + sB)\Big|_{t=s=0},  \quad A = \im[\rho,X],B = \im[\rho,Y].
\]
for some smooth functions $g$ and $g^*$.
\end{theorem}
To prove this theorem we compute the Morozova-Chentsov function for
the metric determined by $g$ and $g^*$ and we get
\[
c(\lambda, \mu) =
\frac{(g(\lambda)-g(\mu))(g^*(\lambda)-g^*(\mu))} {(\lambda-\mu)^2}
\]
From the property $c(t\lambda,t\mu) = t^{-1}c(\lambda,\mu) $ we deduce
that, under the condition $g(0)g^*(0)=0$, $ g(t\lambda)g^*(t\lambda) =
tg(\lambda)g^*(\lambda)$ must hold. This implies
that
\[
g(x)g^*(x)  =  cx    \qquad  (x \in {\bbbr^+} ).
\]
Another necessary condition comes from the property that
$ \lim_{\lambda \to \mu} c(\lambda,\mu) = \mu^{-1} $. In this way, we
arrive at the condition
\[
g^{\prime}(x)g^{*\prime}(x)  = x^{-1}  \qquad (x > 0)
\]
and the equations (24) and (25) together have the solution
$g(x) = ax^p $ and $g^*(x) = bx^{1-p}$,  $ab=c = 1/p(1-p)$,
and the possible limit $\lim_{p \to 0,or 1}$ allowing $x$ and
$\log x$.

\section{Radial extension to pure states}

The idea behind the radial extension comes from the $2\times 2$
case when the Stokes parametrization given by \eqref{eq:Stokes}
identifies $\iM_2$ with the open unit ball in $\bbbr^3$ and the pure
states form the unit sphere. Let us fix a point $P$ in the unit sphere
(i.e. $P$ is a pure state) and a tangent vector $A$ at $P$.
Moreover, let $D$ be an element of the open unit ball except the origin
such that $P$ and $D$ lie on the same radial line $r$. $P$
can be thought as the radial projection of $D$ to the boundary of the
unit ball. Define a tangent vector $\Lift{A}$ at $D$ such that $\Lift{A}$ is orthogonal to {\cal R} and the endpoints
of $A$ and $\Lift{A}$ lie on the same radial line. $\Lift{A}$ can
be thought as a lift of $A$ with respect to the radial
projection. Differential geometers call such lifted vectors
'horizontal vectors' and vectors tangent to the radius at
$D$ are called 'vertical vectors'. Now one can take the inner product
$g_D(\Lift{A},\Lift{B})$ of two lifts $\Lift{A}, \Lift{B}$ of $A, B$
at $D$ with respect to a monotone \Ri metric $g$ and ask for
conditions of the existance of the limit of $g_D(\Lift{A},\Lift{B})$
whenever $D$ goes to $P$ on the radius {\cal R}.

In the general case the radial projection is defined on an open and
dense subset $\iMR_n$ of $\iM_n$ where $\iMR_n$ is formed by the
non-degenerate elements of $\iM_n$, i.e. matrices whose eigenvalues are
all distinct. Now the radial projection $\pi$ is a smooth mapping from
$\iMR_n$ into the pure states $\pure$ such that $\pi(D)$ is the
projection to the one-dimensional eigenspace corresponding to the
largest eigenvalue of $D$. The idea of this projection is that if $D$
is ``near'' to a pure state then the largest eigenvalue of $D$ is near
to 1 and the corresponding eigenspace is one dimensional.

It can be proved that $\iMR_n$ is a fibre bundle over $\pure$ with
projection $\pi$ and in the $2\times 2$ case the fibers are exactly
the radiuses. If $\pi_{*,D}$ denotes the tangent map of $\pi$ at $D$
then the vertical space is $\Ker \pi_{*,D}$ and the horizontal space
$H_D$ is the orthognal complement of $\Ker \pi_{*,D}$ with respect to
a fixed monotone \Ri metric $g$. Since $\pi_{*,D}$ is surjective,
the
restriction of $\pi_{*,D}$ to the horizontal space gives a linear
isomorphism between $H_D$ and the tangent space of $\pure$ at $\pi(D)$
thus for any tangent vector $A$ at $\pi(D)$ there exist a unique lift
$\Lift{A}$ at $D$ such that $\pi_{*,D}(\Lift{A}) = A$.

If $D = \Diag(\lambda_1, \dots, \lambda_n)$ where $\lambda_1$ is the
largest eigenvalue then the vertical vectors at $D$ are identified
with vectors of the following form
\[
      \pmatrix{   x_{11} & 0 & \ldots & 0 \cr
                   0 & x_{22} & \ldots & x_{2n} \cr
                   \vdots & \vdots & \ddots & \vdots \cr
                   0 & x_{n2} & \ldots & x_{nn} \cr
      }
\]
and the horizontal vectors have the form
\begin{equation} \label{eq:HorizontalVector}
    \pmatrix{     0      & \Conj{u}_2    & \ldots & \Conj{u}_n \cr
                  u_2    & 0      & \ldots & 0 \cr
                  \vdots & \vdots & \ddots & \vdots \cr
                  u_n    & 0      & \ldots & 0 \cr
     }.
\end{equation}

The tangent vectors at the pure state $\pi(D) =
\Diag(1,0,\dots,0)$
also have the same form and the lift of a tangent vector is given
by
\begin{equation} \label{eq:Lift}
      \pmatrix{    0 & (\lambda_1-\lambda_2)\Conj{u}_2 & \ldots &
                        (\lambda_1-\lambda_n)\Conj{u}_n \cr
                        (\lambda_1-\lambda_2)u_2 & 0 & \ldots & 0 \cr
                        \vdots & \vdots & \ddots & \vdots \cr
                        (\lambda_1-\lambda_n)u_n & 0 & \ldots & 0 \cr
      }
\end{equation}
which is independent of the choise of $g$.
Now the precise definition of the radial extension is the following

\begin{definition*}
We say that a smooth metric $k$ on $\pure$ is the radial
extension of $g$ if for every $P \in \pure$, for every pair of tangent
vectors $A,\,B$ at $P$ and for every sequence $D_m$ such that
$\pi(D_m) = P$
\[
\lim_{m \to \infty} g_{D_m}(\Lift{A},\Lift{B}) = k_P(A,B).
\]
\end{definition*}
Using \eqref{eq:Lift} one can compute $g_{D}(\Lift{A},\Lift{B})$:
\[
 g_{D}(\Lift{A},\Lift{B}) = 2\Re\sum^n_{i=2} \frac{(\lambda_1 -
        \lambda_i)^2}{f(\lambda_i/\lambda_1)\lambda_1}
        u^i\Conj{v}^i
\]
where $f$ is the operator monotone function corresponding to the
metric and $u_i,\, v_i$ for $i=2,\dots,n$ are the matrix elements of 
horizontal vectors $A,\,B$ as in \eqref{eq:HorizontalVector}. 
Now from this expression it can be easily obtained the
following

\begin{theorem}
Let $g$ be a monotone \Ri metric on $\iM_n$ and let
$f\colon\bbbr^+ \to \bbbr^+$ be the corresponding operator
monotone function.  The radial extension $k$ of $g$ exists if and only
if $f(0) \neq 0$. In this case $k = h/f(0)$ where $h$ is the
canonical \Ri metric on $\pure$, the so called {\it Fubini-Study metric}.
\end{theorem}



\newcommand{\noopsort}[1]{} \newcommand{\printfirst}[2]{#1}
  \newcommand{\singleletter}[1]{#1} \newcommand{\switchargs}[2]{#2#1}
\providecommand{\bysame}{\leavevmode\hbox to3em{\hrulefill}\thinspace}

\end{document}